\documentclass{evn2004}
\usepackage{txfonts}
\usepackage{graphicx}
\begin{document}
   \title{Optical and Radio emission from BL\,Lac objects: long-term trends
          and structural changes}
   \author{E.\,Massaro\inst{1,2}
          \and
          F.\,Mantovani\inst{3}
}
   \institute{ Dipartimento di Fisica, Universit\`a di Roma ``La Sapienza",
              Piazzale A. Moro 2, I-00185 Roma, Italy
         \and
              IASF, Sez. di Roma, CNR-INAF,
              via del Fosso del Cavaliere 100, I-00133 Roma, Italy
         \and
              Istituto di Radioastronomia, CNR,
              Via P. Gobetti, I-40129 Bologna, Italy
             }

   \abstract{
Optical light curves of bright BL Lac objects, in particular those
having the SED synchrotron peak at frequencies lower than 10$^{14}$
Hz, are characterized by long-term trends with time scales of a few decades.
These variations are probabily not related to fast perturbations moving down
the jet, which have a much faster evolution, but more likely to slow changes 
of other physical or geometrical parameters, for example the jet direction.
In these cases we can expect changes of the inner
jet structure detectable with VLBI imaging. Recent literature results
are reviewed and discussed in this context.
   }
\titlerunning{ Optical and Radio emission from BL\,Lac objects}
\maketitle
%

\section{Introduction}

BL Lacertae objects are AGNs characterized by a Spectral Energy
Distribution (SED) having a non-thermal emission over the entire
electromagnetic spectrum.
The typical SED shows two broad peaks, one at lower frequencies
generally explained by Synchrotron Radiation (SR) from relativistic
electrons moving along a jet closely aligned to the observer's
line of sight (Blandford and Rees 1978), and one at higher frequencies
originated by Inverse Compton radiation (ICR) from the same electron
population.
Padovani and Giommi (1995) classified BL Lac objects in two groups
based on the maximum frequency $\nu_S$ of SR peak. The
sources with $\nu_S$ in the range $10^{13} - 10^{15}$ Hz are called
Low-Energy peaked BL Lacs (LBL), those with $\nu_S \simeq
10^{16} - 10^{18}$ Hz are named High-Energy peaked BL Lacs (HBL).

%
   \begin{figure*}
   \centering
   \vspace{307pt}
   \includegraphics{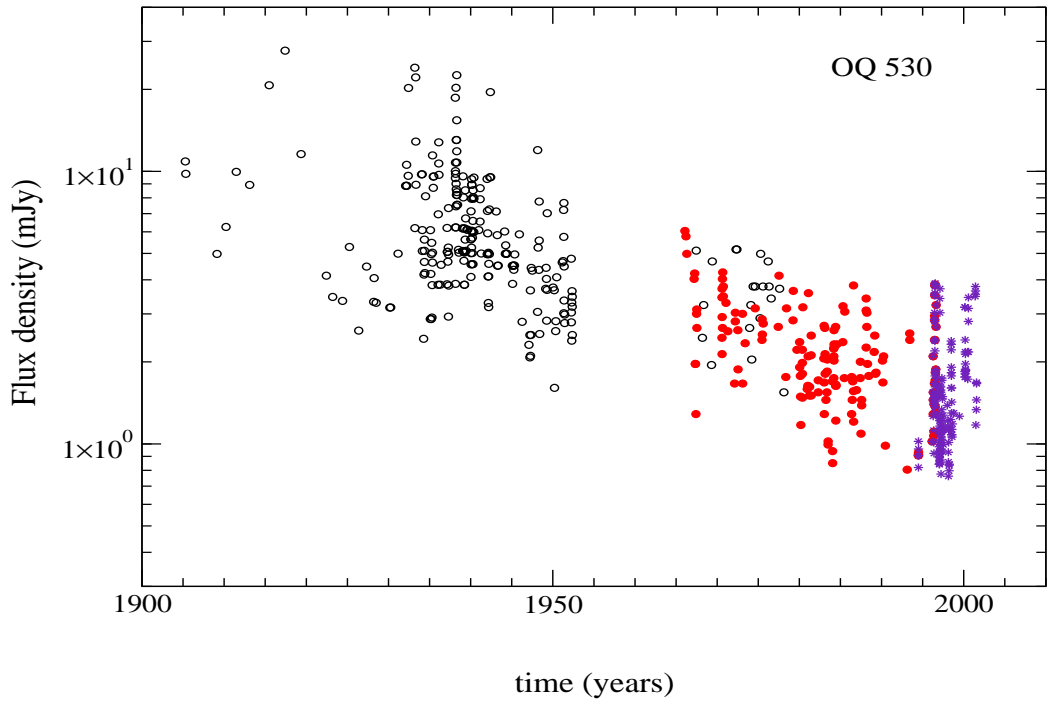}
   \caption{The historic optical light curve of OQ~530 in the B band.
            Open circles are the data from the Harvard archive (Miller 1978)
            corrected to match nearly simultaneous data from Asiago archive
            (filled circles). The most recent points are photometric data
            from (Massaro et al. 2004a).}
               \label{fig:oq530}
   \end{figure*}

LBL sources generally show large variations in their emission
over a wide range of time scales from minutes to years.
Their fast variability, high apparent luminosity and a
brigthness temperature exceeding the Compton limit (Kellermann and
Pauliny-Toth 1969) imply relativistic boosting,
defined by the Doppler beaming factor
$D=1/\Gamma(1-\beta~cos\theta)$,
where $\Gamma=(1-\beta^2)^{-1/2}$ is the bulk Lorentz factor of the
electrons in the jet and $\theta$ the angle to the observer's
line of sight. A good estimate of $D$ is relevant
to understand the physics of BL Lacs. Furthermore, $D$ can change
during the source life time implying large variations of the mean
apparent flux ($S \propto D^{3+\alpha}$).

Optical emission in LBL sources is thought to be the high frequency
tail of the SR component. Therefore, one can expect to observe a correlation
with changes observed in the radio band.
When historic optical light curves are available, it is found that
variations can also occur over time scales of several decades.
Large flux changes with much shorter time scales are
typically superposed to these long term trends. In particular,
the secular variations can be reasonably associated
with changes in the structure and/or direction of the inner jet.
VLBI images are then very useful to obtain informations on the evolution
of the Doppler factor $D$.

In this contribution we present the possible correlation between
changes in the optical luminosity as observed for LBL sources
and the variability observed in their VLBI radio structures.
Many of these sources are also important targets for the
coming space missions for high energy astrophysics
like $AGILE$ and $GLAST$.


\section{Optical historic light curves and monitoring}

Historic light curves of BL Lac objects can be obtained only from
photographic plate archives. They are usually not
well sampled over the maximum time interval of about a century.
An example is the $B$ light curve of OQ~530 shown in Fig. 1 (Massaro
et al. 2004a).
Photographic magnitudes were obtained by Miller (1978) from the
Harvard archive and have been corrected to match the photometric
$B$ band (Nesci and Massaro 1999). Note
the long term trend with a mean decreasing rate of about
15-20\% in ten years.

For the well known source OJ~287 the analysis of the historic data indicated
the existence of recurrent large flares with a period of 11.95 years
(Sillanp\"a\"a et al. 1988) confirmed by the outburst observed in 1994.
A photometric light curve of OJ~287 in the $R$, covering  the past
30 years, is shown in Fig. 2. Note, in addition to the main oubursts of 1982
and 1994, the decreasing trend up to
Spring 1999 when the source luminosity reached its minimum.
After that epoch the brightness of OJ~287 started to increase at a rate
higher than that of the decreasing part. These long term trends have
been indicated in Fig. 2 by the dashed tick line drawn following the local
minima. Note also that the highest flux level reached by OJ~287 in
the outburst of 1994 is about one third of that of 1982 maximum
and seems roughly proportional to the mean flux.
If the periodic recurrence of these oubursts will be
confirmed in future, the next one, expected for 2006, could reach a flux
comparable or even higher than that of 1982.

Another well studied source is ON~231 (Massaro et al. 2001).
Its behaviour was characterised by an increase of the mean brightness
after a minimum in the early seventies culminated in a very strong
outburst in April-May 1998 (Massaro et al. 1999). After that event
the mean source luminosity slowly declined, continuing to show
changes greater than one magnitude (Tosti et al. 2002).

The amount of available data on variability of
bright LBL objects has largely increased in the past ten years
thanks to the use of small aperture
telescopes equipped with CCD detectors.
It is now simple to make these instruments fully automatic obtaining a very
high observational efficency. Automatic optical monitoring provided very
well sampled light curves for several sources (see for instance
http://astro.fisica.unipg.it/PGblazar/tabella2000.htm). They provide prompt
information on the onset of active phases, important to trigger observations
from space X- and $\gamma$-ray observatories.
The best known case is that of BL Lac itself which, after a
rather quiescent phase, in 1997 showed a very intense activity characterized by
a very high luminosity and a large intraday variability.

   \begin{figure*}
   \centering
   \vspace{307pt}
   \includegraphics{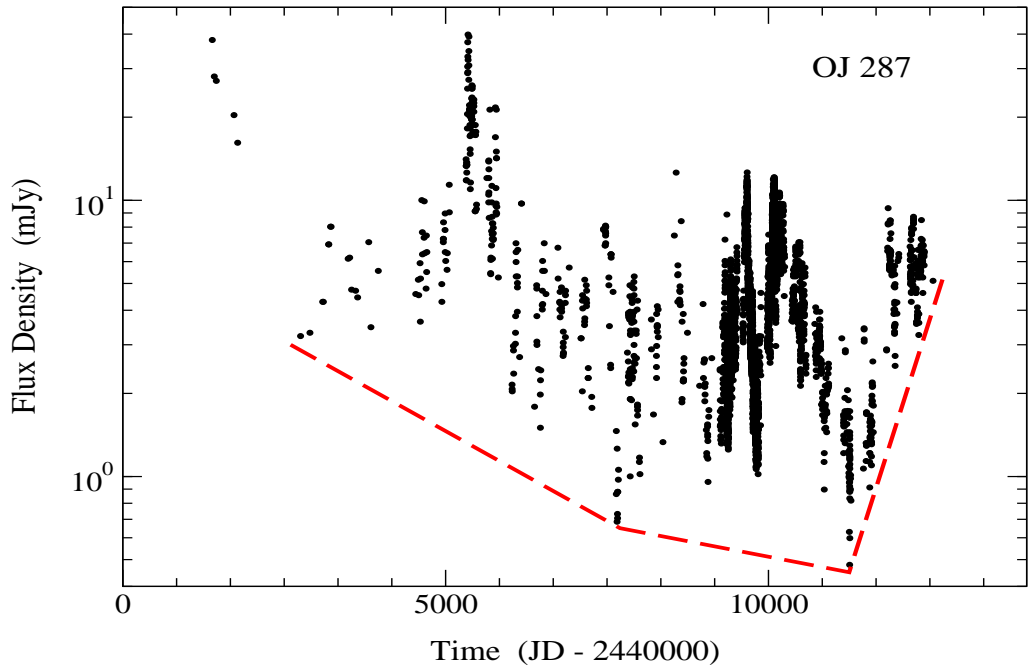}
   \caption{The photometric $R$ light curve of OJ~287 in the R band
            (not corrected for the interstellar reddening).
            The two outbursts of 1982 (around JD= 5,400) and 1994
            (around JD=9,600) are evident. Note the mean decresing
            trend (dashed line) down to the minimum observed in March 1999
            followed by a faster increase continuing up to now.
            (Thanks to N. Marchili and G. Tosti).}
         \label{fig:oj287}
    \end{figure*}

In the radio band LBL objects are also variable but generally not so
violently variable as in the optical. Monitoring observations in the GHz
range are important to search for correlations between optical and IR data,
useful to investigate
the radiative life times of relativistic electrons. The most continuous
radio monitoring programs have been carried out for several years by
the groups of the University of Michigan
(http://www.astro.lsa.umich.edu/obs/radiotel/radiotel.html) and of
the Metsah\"ovi Radio Observatory (http://kurp-www.hut.fi/quasar/;
Ciaramella et al. 2004).

\section{The Spectral Energy Distribution}

To better understand the link between the optical and the radio emission in LBL
sources it is important to have a good description of the broad band
spectral properties of the SR. Non-thermal spectra are generally well
fitted by single power laws over frequency intervals of the order of one
decade. However, on wider intervals power law models fail to reproduce 
the observed mild curvature.
Landau et al. (1986) showed that the spectra of some BL Lac objects
from radio to the optical and UV ranges are represented by a
parabola in lograrithmic coordinates

\begin{equation}
S(\nu) = A (\nu/\nu_o)^{-(a + b Log(\nu/\nu_o))}
\end{equation}

\noindent
better than an exponential cut-off. Here the SED curvature is described
only by the parameter $b$.
This spectral law has been used successfully to fit the X-ray SEDs for
the HBL sources Mkn~421 and Mkn~501 (Massaro et al. 2004b,c)
in a variety of luminosity states. This implies
that the electron spectra also show a similar distribution, which
can be obtained from statistical acceleration mechanisms when the
probability for a particle to remain inside the acceleration region
decreases with the energy of the particle itself.
Alternatively, when the particle energy remains below a critical value
the acceleration probability is constant and therefore the resulting
spectrum is given by a power law and changes into a log-parabola when
the accleration probabilty decreases.
The resulting SR spectrum can be then approximated by

\begin{equation}
S(\nu) = A \nu^{-(a + b Log(1+\nu/\nu_1))}~~~.
\end{equation}

Some spectra of the SR observed from LBL objects are plotted in
the upper panel of Fig. 3.
Equation (2) has been used for the two spectra of
OQ~530, while Equation (1) for the remaining sources.

Another interesting possibility is that presented in Fig. 3 (lower panel)
in which
there are two SR components peaking at different frequencies.
The two spectra were computed using the same values of $b$ as the electron 
populations were accelerated by similar processes. 
They could be originated in different regions along the jet and evolve with
different time scales. Simultaneous broad 
band observations may not provide enough information to distinguish the 
two components, while variability studies can give the most useful information.

%
   \begin{figure}
   \centering
   \vspace{325pt}
   \includegraphics{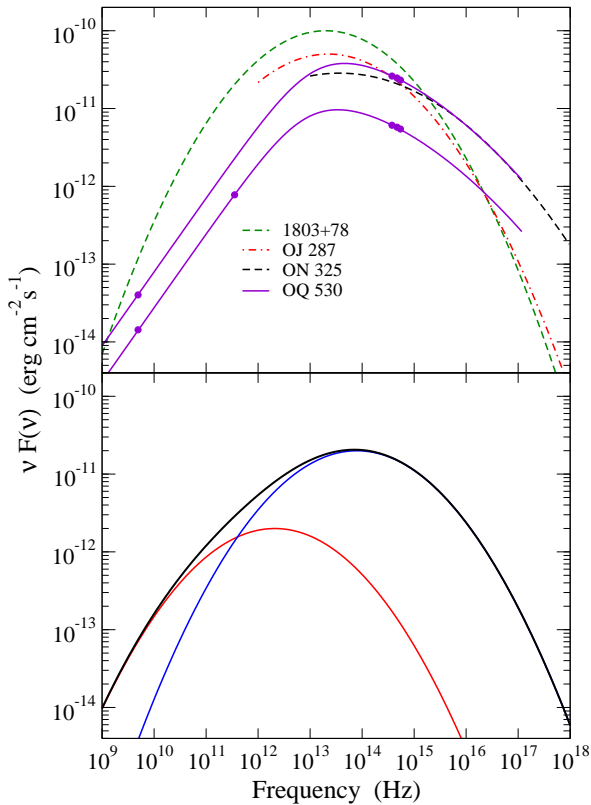}
   \caption{Upper panel: Spectral Energy Distribution of some LBL objects. 
            (adapted from Perri et al. 2003 and Massaro et al. 2004a).
            Lower panel: a possible model of a two component SED. }
               \label{fig:sedsed}
   \end{figure}
%
%


\section{VLBI imaging}

VLBI imaging of BL Lac objects is important because we can obtain
direct information on the evolution of the inner jet structure.
Furthermore, polarisation measurements allow us to know the orientation 
of the magnetic field (see for instance Gabuzda et al. 1999, 2004). 
At the typical distances of bright BL Lacs objects,
say $z \lesssim 0.3$, the VLBI angular resolution at 5 GHz corresponds
to distances of few parsecs. As a consequence, only structure 
changes involving time scales of years can be detected.
Images at higher frequencies (say 43\,GHz) are very useful to investigate the
variations on shorter time scales. However, observations at lower frequencies
are needed to define the spectral behaviour.
Savolainen et al. (2002) applied a model for the time evolution
of radio flares of a blazar sample, including some LBL objects,
observed at 22 and 43\,GHz. Their analysis was successful in describing
events with typical time scales of one-two years. They showed that in more
than 50\% of the sources, a new VLBI 
component appeared in the jet after a total flux density flare.
Usually these components brighten and decay within
$\sim$0.15 mas of the radio core. This result supports the 
interpretation that flares are originated by a sudden energy 
injection from the core. Subsequent shocks propagate
in the jet and are associated with superluminal components.
For faster variations, like those frequently observed in the optical for LBL 
sources which evolve on time scale of a month or even shorter, 
new components emerging from the core cannot be easily seen. 
They are likely exhausted before they can reach an angular 
separation large enough to be resolved out by VLBI. 

Long term optical trends are likely originated by
different and more regular processes. An interesting possibility
is that of a precessing jet, which implies a change of $\theta$
and consequently of the Doppler factor $D$. One
can expect to observe a relation between the superluminal
velocity of the new born components and the observed total core luminosity.
This relation is shown in Fig. 4 where the Doppler factor is plotted
as a function of $\beta_{app}$ (Blandford and K\"onigl 1979). For 
$\theta$ smaller than about 8$^\circ$, 
an increase of $\theta$ would imply a higher superluminal velocity 
corresponding to a fading of the luminosity and $vice versa$. 
Of course, it is not easy
to separate such an effect from those due to short time changes.
Dense radio-optical and VLBI monitorings (typically twice a
year) are here required.

Evidence for a precessing jet nozzle in BL Lacertae has been recently
found by Stirling et al. (2003) using a series of VLBA high resolution
and polarisation images at 43 GHz taken from March 1998 to April 2001. 
They estimated a period of $\sim$2 years. There is a very large amount
of optical data on this source, in particular after the large
outburst of 1997, however, a complete analysis has not been carried out.
Fan et al. (1998) analysed the historic light curve of BL Lacertae
and found several recurrence times (one of them of about 2 years).
Further investigations to confirm this result, in particular on the phase 
of the periodic changes, are necessary. Furthermore, this
analysis would be useful to understand if the spectrum of the SR
observed in the optical extends to the radio band or if it is generated
by a different component. 

%
   \begin{figure}
   \centering
   \vspace{240pt}
   \includegraphics{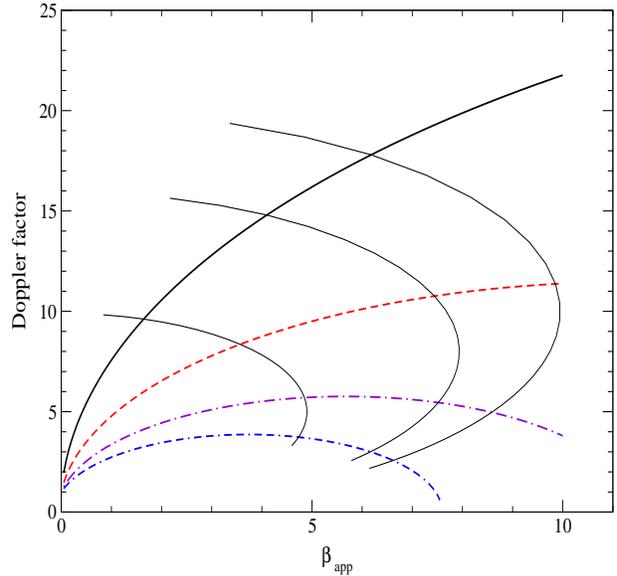}
   \caption{The Doppler factor as a function of the apparent velocity 
           (in units of the speed of light) for $\theta$ = 2$^{\circ}$, 
           5$^{\circ}$, 10$^{\circ}$, 15$^{\circ}$ (from top to bottom).
           The other lines show the relation between $D$ and $\beta_{app}$
           when $\theta$ is changing and $\Gamma$ is taken equal to 
           5, 8 and 10.
              }
               \label{fig:dopfbe}
   \end{figure}

EVN images at 5GHz of OQ~530, recently presented by Massaro
et al. (2004a), show a bright radio core and a rather weak jet.
Optical and radio light curves are charaterized by a low
luminosity phase at the beginning of 1999 and they start to 
brighten later on. Simultaneous optical and radio 
observations (February 1999 and June 2001) indicate a  
similar increase of the core flux in the $R$ band and at 5 GHz
(Fig. 3 in Massaro et al. 2004a) consistent with a unique SR emission 
component.

An interesting point to rise discussing the structure of BL Lac objects 
(and more in general for blazars) is the presence of stationary components 
detected near the core.
In the extensive VLBI monitoring at 22 and 43 GHz on a sample of 42 
$\gamma$-ray loud blazars, including 11 LBL objects Jorstad et al. 
(2001) and Marscher et al. (2002) found stationary components
within a distance of 2 mas from the core in 27 sources.
They proposed that those components could be associated with a
standing hydrodynamical compression. In such a case some components
may appear blended in a single feature without a significant
proper motion making it difficult to detect superluminal motion in
the inner part of the jet.

Finally it is worth to mention the case of ON\,231 (Massaro et al. 2001;
Mantovani \& Massaro, this conference) were a component not previously 
detected was observed on the opposite side of the jet with respect to
the radio core. Again, the change of the direction of
the jet, occurring after the Spring 1998 optical flare, was considered
as a possible explanation for such unusual structure.

\section{Conclusions }

The next ten years will be very interesting for the astrophysics of BL Lac
objects. These sources and Flat Spectrum Radio Quasars, grouped 
together in the Blazar class, are charaterized by a strong and variable
non-thermal emission. For this reason they are the most numerous
extragalactic $\gamma$-ray sources and are also well detected in the
IR range. The majority of the 208 sources detected by $WMAP$ (Bennett et al. 
2003) are indeed blazars (Giommi \& Colafrancesco 2004).

Two high sensitivity space observatories, well 
suited to observe BL Lac objects, will be operative in the next three 
years: $GLAST$ for $\gamma$ rays above $\sim$30 MeV and $Planck$ 
(30\,GHz -- 857\,GHz). It is expected that at least a few thousands
sources will be detected. It is important at this stage to select
a sample of target sources planning a monitoring programme both in the
radio and optical bands.
In particular, it will be very useful to plan for a monitoring programme
with high resolution radio interferometers to obtain detailed information on
milli-arcsecond scale structural changes and brightness variations
of the broad-band synchrotron emission.
  
Optical variability plus VLBI imaging can also be useful to 
verify the unified scheme model for BL Lac objects. According to that 
interpretation, the relativistic effects are relevant when the angles 
between the jet direction and the observer's line of sight are small.
When this angle increases the source becomes fainter, 
its variability time scale becomes longer and, at large angles the
source should appear as a FRI radiogalaxy (Ghisellini et al. 1993).
Some BL Lac objects are rather faint in the optical while their
radio flux is relatively high. The synchrotron peak frequency 
of these sources is rather low, $\nu_S < 10^{13}$ Hz. We will 
refer to them as Very Low energy peaked BL Lac objects,
(shortly VLBL). 
In the unification scenario VLBL objects could have
Doppler factors smaller than that of typical LBL sources. Their
behaviour should be intermediate between these sources and FRI radio
galaxies. Unfortunately, too few data and images are available for
this 'class' of objects. Therefore no firm conclusions
can be at present derived. 
It will be then useful to plan coordinated observational programs 
for a sample of VLBL objects to investigate their main properties.
In particular, simultaneous broad band observations are needed
to have information on the SED and VLBI observations the detect 
superluminal motions of the jet components to estimate the Doppler factor.

%
%

\begin{acknowledgements}
We are grateful to Paolo Giommi, Roberto Nesci, Matteo Perri and Gino 
Tosti for useful comments and discussions.
\end{acknowledgements}

\cleardoublepage


\begin{thebibliography}{ }
\small

\bibitem [2003]{bennett03} Bennett C.L. et al. 2003, ApJS 143, 97

\bibitem [1978]{blanrees78} Blandford R.D., Rees M.J. 1978, Phys. Scripta 17, 265

\bibitem [1979]{blankon79} Blandford R.D., K\"onigl A. 1979, ApJ 232, 34

\bibitem[2004]{ciaram04} Ciaramella A., Bongardo C. et al. 2004, A\&A 419, 485

\bibitem[1998]{fan98} Fan,J.H., Xie G.Z. et al. 1998, ApJ 507, 173

\bibitem[1999]{gabuzda99} Gabuzda D.C., Pushkarev A.B., Cawthorne T.V.  1999, MNRAS 307, 725

\bibitem[2004]{gabuzda04} Gabuzda D.C., Murray E., Cronin P. 2004, MNRAS 351, 89

\bibitem[1993]{ghisel93} Ghisellini G., Padovani P. et al. 1993, ApJ 407, 65

\bibitem[2004]{giocol04} Giommi P., Colafrancesco S. 2004, A\&A 414, 7

\bibitem[2001]{jorstad01} Jorstad S.G., Marscher A.P. et al. 2001, ApJS 134, 181

\bibitem [1969]{kellpaul69} Kellerman K.I., Pauliny-Toth I.I.K. 1969, ApJ 155,
L71

\bibitem[1986]{landau86} Landau R., Golisch B. et al. 1986, ApJ 348, 14

\bibitem[2002]{marscher02} Marscher A.P., Jorstad S.G. et al. 2002, ApJ 577, 85 

\bibitem[1999]{massaro99} Massaro E., Maesano M. et al. 1999, A\&A 342, L59

\bibitem[2001]{massaro01} Massaro E., Mantovani F. et al. 2001, A\&A 374, 435

\bibitem[2004]{massaro04} Massaro E., Mantovani F. et al. 2004a, A\&A 423, 935

\bibitem[2004]{massaro04} Massaro E., Perri M. et al. 2004b, A\&A 413, 489

\bibitem[2004]{massaro04} Massaro E., Perri M. et al. 2004c, A\&A 422, 103

\bibitem[1978]{miller78} Miller H.R. 1978, ApJ 223, L67 

\bibitem[1999]{nesci99} Nesci R., Massaro E. 1999 Proc. "Treasure-Hunting in
 Astronomical Plate Archives", (P. Kroll, C. La Dous, H.-J. Brauer eds.), 
 Sonneberg,p. 111

\bibitem[1995]{padov95} Padovani P., Giommi P. 1995, ApJ 444, 567

\bibitem[2003]{perri03} Perri M., Massaro E. et al. 2003, A\&A 407, 453

\bibitem[2002]{savol02} Savolainen T., Wiik K. et al. 2002 A\&A 394, 851

\bibitem[1988]{sillan88} Sillanp\"a\"a A., Haarala S. et al. 1988, ApJ 325, 635

\bibitem[2003]{stirling03} Stirling A.M., Cawthorne T.V. et al. 2003, MNRAS 341, 405

\bibitem[2002]{tosti02} Tosti G., Massaro E. et al. 2002, A\&A 395, 11

\end{thebibliography}
\end{document}